\documentclass[prd,twocolumn,superscriptaddress,showpacs,preprintnumbers,bibnotes,nofootinbib,floatfix]{revtex4}

\usepackage{dcolumn}
\usepackage{amsmath,amssymb,graphicx}

\def\a{\alpha}
\def\d{\delta}
\def\g{\gamma}
\def\ve{\varepsilon}
\def\th{\theta}
\def\k{\kappa}
\def\l{\lambda}
\def\s{\sigma}
\def\o{\omega}

\def\D{\Delta}
\def\L{\Lambda}
\def\O{\Omega}

\def\hs{\hspace}
\def\ol{\overline}
\def\no{\nonumber}

\def\lf{\left}
\def\rg{\right}
\def\la{\langle}
\def\ra{\rangle}

\def\be{\begin{equation}}
\def\ee{\end{equation}}

\def\td{\tilde}
\def\+{\text{\raisebox{0.3ex}{$\scriptstyle +$}}}
\def\-{\text{\raisebox{0.3ex}{$\scriptstyle -$}}}

\newlength{\nl}
\newcommand{\iol}[1]{%
\settowidth{\nl}{$#1$}%
\shortstack{\rule{0.8\nl}{0.15ex}\\[-0.5ex]$#1$}}

\begin{document}

\preprint{ADP-06-05/T636, JLAB-THY-06-496}

\title{EMC and Polarized EMC Effects in Nuclei}

\author{I.~C.~Clo\"et}
\email{icloet@jlab.org}
\affiliation{Special Research Centre for the Subatomic Structure of Matter and \\
             Department of Physics and Mathematical Physics, University of Adelaide,
             SA 5005, Australia}
\affiliation{Jefferson Lab, 12000 Jefferson Avenue, Newport News, VA 23606, U.S.A.}
\author{W.~Bentz}
\email{bentz@keyaki.cc.u-tokai.ac.jp}
\affiliation{Department of Physics, School of Science, Tokai University, 
                         Hiratsuka-shi, Kanagawa 259-1292, Japan}
\author{A.~W.~Thomas}
\email{awthomas@jlab.org}
\affiliation{Jefferson Lab, 12000 Jefferson Avenue, Newport News, VA 23606, U.S.A.}

\begin{abstract}
We determine nuclear structure functions and quark distributions for $^7$Li, $^{11}$B,
$^{15}$N and $^{27}$Al.   
For the nucleon bound state we solve the covariant quark-diquark equations in
a confining Nambu--Jona-Lasinio model, which yields excellent results for the 
free nucleon structure functions. 
The nucleus is described using a relativistic shell model, including
mean scalar and vector fields that couple to the quarks in the nucleon.
The nuclear structure functions are then 
obtained as a convolution of the structure function of the bound nucleon with the 
light-cone nucleon distributions. 
We find that we are readily able to reproduce the EMC effect in finite nuclei and 
confirm earlier nuclear matter studies that found a large polarized EMC effect. 
\end{abstract}

\maketitle

%===============================================================================
%###############################################################################
%===============================================================================
\section{Introduction}

One of the greatest challenges confronting nuclear physics is to understand 
how the fundamental degrees of freedom -- the quarks and gluons -- give rise to
the nucleons and to inter-nucleon forces that bind nuclei. Quark models such as 
the quark-meson coupling model (QMC)~\cite{Guichon:1987jp,Saito:1995up,Guichon:1995ue} in which the 
structure of the nucleon is self-consistently modified by the nuclear medium, 
can be re-expressed in terms of local effective forces which closely resemble the 
widely used and successful Skyrme forces~\cite{Guichon:2004xg,Guichon:2006er}. 
While this opens the 
possibility to describe the low energy nuclear structure in terms 
of quark degrees of freedom, it is also important to identify phenomena 
which provide explicit windows into quark-gluon effects in nuclei.
Probably the most famous candidate
is the EMC effect \cite{Aubert:1983xm,Arneodo:1992wf,Geesaman:1995yd}, 
which refers to the substantial 
depletion of the in-medium spin-independent nucleon 
structure functions in the valence 
quark region, relative to the free structure functions.

Considerable experimental and theoretical effort has been invested to try to
understand the dynamical mechanisms responsible for the EMC effect. It is now 
widely accepted that binding corrections at the nucleon level cannot account 
for the observed 
depletion and a change in the internal structure of the nucleon-like
quark clusters in nuclei is required~\cite{Piller:1999wx,Miller:2001tg,Cloet:2005rt}. 
Although the EMC effect has received the 
most attention, there are a number of other phenomena which may require
a resolution at the quark level, such as the quenching of spin matrix elements 
in nuclei \cite{Arima:1988xa} and the quenching of the 
Coulomb sum-rule~\cite{Morgenstern:2001jt,Aste:2005wc}. 
Important hints for medium modification also come from 
recent electromagnetic form factor measurements on $^4$He 
\cite{Strauch:2002wu,Dieterich:2000mu}, which  suggest a reduction of the 
proton's electric to magnetic form factor ratio in-medium.
Sophisticated nuclear structure calculations fail to fully account for the 
observed effect \cite{Udias:2000ig} and agreement with the
data is only achieved by
also including a small change in the internal structure
of the nucleon \cite{Dieterich:2000mu}, predicted a number of years 
before the experiment~\cite{Lu:1997mu}.

The focus of this work is on the medium modifications to the nucleon structure
functions in nuclei. We calculate the nuclear quark distributions explicitly
from the quark level using the convolution formalism \cite{Jaffe:1985je}. 
The quark distributions in the bound nucleon 
are obtained using a confining 
Nambu--Jona-Lasinio (NJL) model, where the nucleon is described as a 
quark-diquark bound state in the relativistic Faddeev formalism.
The nucleon distributions in the nucleus are determined using a 
relativistic single particle shell model, including scalar and
vector mean-fields that couple to the quarks in the nucleon.
This model, which is very similar in spirit to the QMC model, has the advantage 
that it is completely covariant, so that one can apply standard field 
theoretic methods to the calculation of the structure functions. 
Using this approach we are readily able to reproduce the EMC effect in nuclei.
However, the main focus of this paper is on a new ratio -- 
the nuclear spin structure function, $g_{1A}$, divided by 
the naive free result -- which we refer to as the ``polarized EMC effect''. 

%===============================================================================
%###############################################################################
%===============================================================================
%\section{Deep inelastic scattering from nuclear targets}

The formalism to describe deep inelastic scattering (DIS) from 
a target of arbitrary spin was developed 
in Refs.~\cite{Hoodbhoy:1988am,Jaffe:1988up}. 
We focus on results specific to the Bjorken limit, expanding on 
those points necessary for the following discussion.

When parameterized in terms of structure functions, the hadronic tensor 
in the Bjorken limit has the form
\begin{multline}
W_{\mu\nu}^{JH} = \lf(g_{\mu\nu} \frac{P\cdot q}{q^2}\, +\, \frac{P_\mu P_\nu}
{P\cdot q} \rg)\, F_{2A}^{J H}(x_A)\\
+ i\,\frac{\ve_{\mu\nu\l\s}q^\l P^\s}{P\cdot q}\, g_{1A}^{J H}(x_A),
\label{eqn:hadtensor}
\end{multline}
for a target of 4-momentum $P^\mu$, total angular momentum $J$ and helicity $H$ 
along the direction of the incoming electron momentum.  
In obtaining Eq.~(\ref{eqn:hadtensor}) 
we have used a generalization of the Callen-Gross relation, $F_{2A}^{JH} = 2\,\hat{x}_A\,F_{1A}^{JH}$,
and ignore terms proportional to $q_\mu$ as 
the lepton tensor is conserved.
We define the Bjorken scaling variable as 
\be
x_A = A\,\hat{x}_A =A\,\frac{Q^2}{2\,P\cdot q},
\label{eqn:xA}
\ee
so that the structure functions have support in the domain
$0 < x_A \leqslant A$.

In the Bjorken limit the nuclear structure functions can be 
expressed as
\begin{align}
\label{eqn:FJH}
F_{2A}^{JH}(x_A) &= \sum_q e_q^2\, x_A\, \lf[q_{A}^{JH}(x_A) + \ol{q}_{A}^{JH}(x_A)\rg], \\
\label{eqn:gJH}
g_{1A}^{JH}(x_A) &= \frac{1}{2}\sum_q e_q^2 \lf[\D q_{A}^{JH}(x_A) + \D \ol{q}_{A}^{JH}(x_A)\rg],
\end{align}
where $q$ represents the flavour and
\begin{align}
q^{JH}_A(x_A) &= q^{JH}_{A\uparrow}(x_A) + q^{JH}_{A\downarrow}(x_A),\\
\D q^{JH}_A(x_A) &= q^{JH}_{A\uparrow}(x_A) - q^{JH}_{A\downarrow}(x_A),
\end{align}
are generalizations of the usual spin-$\tfrac{1}{2}$ quark distributions.
The quark distributions, $q^{JH}_{As}(x_A)$, are interpreted as:
\textit{the probability to find a quark (of flavour $q$) with lightcone
momentum fraction $x_A/A$ and spin-component $s$ in a target with helicity $H$}.
Parity invariance of the strong interaction requires 
$q_{As}^{J H} = q_{A\,\-\!s}^{J\, \-\!H}$,
so that $F_{2A}^{JH} = F_{2A}^{J\,\-\!H}$
and $g_{1A}^{JH} = -g_{1A}^{J\,\-\!H}$ and hence there are $2J+1$ 
independent structure functions for a spin-$J$ target. 

For DIS on targets with $J \geqslant 1$ it is more convenient to work with
multipole structure functions or quark distributions \cite{Jaffe:1988up} 
rather than the helicity dependent quantities discussed above. The 
helicity and multipole representations are related by the following transformations
\begin{align}
\label{eqn:FJK}
F_{2A}^{(JK)} &= \sum_{H = \-\!J,\ldots,J}\, A^{JK}_H\,F_{2A}^{JH}, \quad K=0,2,\ldots,2J, \\
\label{eqn:gJK}
g_{1A}^{(JK)} &= \sum_{H = \-\!J,\ldots,J}\, A^{JK}_H\,g_{1A}^{JH}, \quad K=1,3,\ldots,2J,
\end{align}
where 
\be
A^{JK}_H = \lf(-1\rg)^{J-H}\sqrt{2K+1}
\begin{pmatrix} J&J&K \\ H&-H&0 \end{pmatrix},
\ee
and $\lf(\cdots\rg)$ is a Wigner $3j$-symbol.
Identical multipole expansions can also be defined for the spin-independent
and spin-dependent quark distributions. Comparing the inverse of these 
relations with the familiar Wigner-Eckart theorem, it is clear that
$q_A^{(JK)}$ and $\D q_A^{(JK)}$ are reduced matrix elements of multipole
operators of rank $K$.

For nuclear targets the multipole formalism has several 
advantages, these include
\begin{itemize}
\item $F_{2A}^{(J0)} = \sqrt{2J+1}\,F_{2A}$, where $F_{2A}$ is
the familiar spin-averaged structure function.
\item The number and spin sum-rules are completely saturated by 
the lowest multipoles, $K=0$ and $K=1$ respectively.
\item In a single particle (shell) model for the nucleus, the spin saturated
core contributes only to the $K=0$ multipole and all $K>0$ contributions
come from the valence nucleons.
\item In all cases investigated in this paper, we find that the lowest
multipoles, $K=0$ for spin-independent and $K=1$ for spin-dependent,
are by far the dominant distributions.
\end{itemize}

%===============================================================================
%###############################################################################
%===============================================================================
\section{Nuclear distribution functions}

The twist-2, spin-dependent quark distribution in a nucleus of mass number $A$,
momentum $P^{\mu}$ and helicity $H$ is defined as
\begin{multline}
\D q_A^{J H}\lf(x_A\rg) = \frac{P_-}{A} \int\frac{d\o^-}{2\pi} e^{iP_-\,x_A\,\o^-/A} \\
\la A,P,H \lvert \ol{\psi}_q(0)\,\g^+\g_5\,\psi_q(\o^-)\rvert A,P,H \ra,
\label{eqn:nucdis}
\end{multline}
where $\psi_q$ is the quark field. To evaluate Eq.~(\ref{eqn:nucdis})
we express it as the convolution of a quark distribution
in a bound nucleon, with the nucleon distribution
in the nucleus \cite{Jaffe:1985je}. 
If a shell model is used to determine the nucleon 
distribution, then in the convolution formalism 
Eq.~(\ref{eqn:nucdis}) has the form 
\begin{align}
\D q_A^{J H}\lf(x_A\rg) &= \sum_{\a,\k m}\, C^{JH}_{\a,\k m}
\int_0^A\!\! dy_A\!\! \int_0^1\!\! dx \no \\
&\hs{10mm} \d\!\lf(x_A-y_A\, x\rg) \D q_{\a,\k}\!\lf(x\rg)\ \D f_{\k m}\!\lf(y_A\rg), \no \\
&\equiv \sum_{\a,\k m} C^{JH}_{\a,\k m}\, \D q_{\a,\k}^m\!\lf(x_A\rg),
\label{eqn:con}
\end{align}
where $\a \in (p,n)$ label the nucleons and the sum over the Dirac quantum number $\k$ and $j_z=m$ 
(that is, the occupied single particle states)
is such that the coefficients $C^{JH}_{\a,\k m}$ guarantee the 
correct quantum numbers $J$, $H$, $T$ and $T_z$ for the nucleus.
Note, in Eq.~(\ref{eqn:con}) a sum over the principle quantum number
$n$ is implicit.

The function $\D f_{\k m}(y_A)$ is the spin-dependent nucleon distribution 
(in the state $\k m$) in the nucleus and is given by
\begin{multline}
\D f_{\k m}(y_A) = \sqrt{2} \int \frac{d^3p}{\lf(2\pi\rg)^3} 
\ \d\! \lf(y_A - \frac{p^3 + \ve_\k}{\iol{M}_N}\rg) \times \\
\iol{\Psi}_{\k m}\!\lf(\vec{p}\,\rg)\,\g^+\g_5\,\Psi_{\k m}\!\lf(\vec{p}\,\rg),
\label{eqn:gyA}
\end{multline}
where $\ve_\k$ is the single particle energy, $\Psi_{\k m}(\vec{p}\,)$ 
are the single particle Dirac wavefunctions in momentum space and $\iol{M}_N = M_A/A$ 
is the mass per nucleon. 
Implicit in our definition of the convolution formalism used in Eq.~(\ref{eqn:con}) 
is that the quark distributions in the bound nucleon, 
$\D q_{\a,\k}(x_A)$, respond 
to the nuclear environment. Expressions for the
spin-independent distributions are obtained by simply 
replacing $\g^+\g_5$ with $\g^+$.

First we obtain expressions for the nucleon distributions in the nucleus.
The central potential Dirac eigenfunctions have the general form 
\be
\Psi_{\k m}(\vec{p}\,) = 
\lf(-i\rg)^\ell\begin{bmatrix} 
F_\k(p)\ \O_{\k m}\!\lf(\th,\phi\rg) \\
~G_\k(p)\  \O_{\-\!\k m}\!\lf(\th,\phi\rg)
\end{bmatrix},
\label{eqn:Dwave}
\ee
where $F_\k$ and $G_\k$ are the radial wavefunctions in momentum 
space and $\O_{\k m}$ are the spherical two-spinors. 

Substituting Eq.~(\ref{eqn:Dwave}) into Eq.~(\ref{eqn:gyA}) and also the spin-independent
equivalent we obtain the following expressions for the single nucleon $k$-multipole distributions in 
the nucleus
\vskip -1.5em
\begin{widetext}
\begin{align}
f_{\k k}(y_A) &= (-1)^{j+\frac{1}{2}}\lf(2j+1\rg)\lf(2\ell+1\rg)\sqrt{2k+1}\,
\begin{pmatrix} \ell&k&\ell\\ 0&0&0 \end{pmatrix}
\begin{Bmatrix} \ell & k & \ell \\ j & s & j \end{Bmatrix} \no \\
&\hs{25mm} \frac{\iol{M}_N}{16\pi^3} \int_\L^\infty \, dp\ p \, 
\lf[F_\k(p)^2 + G_\k(p)^2 + \frac{2}{p}\lf(\ve_k-\iol{M}_N\,y_A\rg)F_\k(p)G_\k(p) \rg]\,
P_k\!\lf(\frac{\iol{M}_N\,y_A-\ve_\l}{p}\rg),
\label{eqn:fkk}
\end{align}
\vskip -0.5em
\begin{align}
\D f_{\k k}(y_A) &= \lf(2j+1\rg)\sqrt{2k+1}\frac{\iol{M}_N}{16 \pi^3} \int_\L^\infty dp\, p\no \\
&\lf\{2\,P_k\!\lf(\frac{\iol{M}_N\,y_A - \ve_\l}{p}\rg)
F_\k(p) G_\k(p)(-1)^{j - \frac{1}{2}} \sqrt{(2\ell+1)(2\td{\ell}+1)}
\begin{pmatrix} \ell&k&\td{\ell}\,\\ 0&0&0\, \end{pmatrix}
\begin{Bmatrix} \ell & k & \td{\ell} \\ j & s & j \end{Bmatrix} 
\begin{matrix} \\ \\ \\ \end{matrix} \rg. \no \\
&\hs{10mm}-\sqrt{6} \lf(-1\rg)^{\ell}\sum_{L=k-1,k+1} \lf(2L+1\rg)
\,P_L\!\lf(\frac{\iol{M}_N\,y_A - \ve_\l}{p}\rg)\no
\begin{pmatrix} L&1&k \\ 0&0&0 \end{pmatrix} \no \\
&\hs{27mm}\lf.\lf[F_\k(p)^2 (2\ell+1)\begin{pmatrix} \ell&L&\ell \\ 0&0&0 \end{pmatrix}
\begin{Bmatrix} \ell & s & j \\ L & 1 & k \\ \ell & s & j \end{Bmatrix}
- G_\k(p)^2 (2\td{\ell}+1) \begin{pmatrix} \td{\ell}&L&\td{\ell} \\ 0&0&0 \end{pmatrix}
\begin{Bmatrix} \td{\ell} & s & j \\ L & 1 & k \\ \td{\ell} & s & j \end{Bmatrix}
\rg]
\begin{matrix} \\ \\ \\\end{matrix}\rg\},
\label{eqn:gkk}
\end{align}
\end{widetext}
where $P_k$ are Legendre polynomials of degree $k$ and $\L=\lf|\iol{M}_N\,y_A-\ve_\k\rg|$.
In deriving Eq.~(\ref{eqn:fkk}) it is convenient to use the identity 
$\O_{\-\!\k m} = - \lf(\vec{\s}\cdot\hat{p}\rg)\,\O_{\k m}$.

The single nucleon wavefunctions (Eq.~(\ref{eqn:Dwave})) are solutions of the Dirac
equation with scalar, $S_N(r)$, and vector, $V_N(r)$, mean-fields. In principle 
these fields should be calculated self-consistently in our (NJL) 
model framework by minimizing the
total energy of the system, as was done in Refs.~\cite{Mineo:2003vc,Cloet:2005rt}
for nuclear matter. Instead we choose Woods-Saxon
potentials for $S_N(r)$ and $V_N(r)$.
The depth parameter of each potential is set 
to the strength of the scalar or vector field obtained
from an earlier self-consistent nuclear matter calculation, that is
$S_0 = -194~$MeV and $V_0 = 133~$MeV \cite{Cloet:2005rt} 
and we choose standard  values for the range $R=1.2~A^{1/3}~$fm and 
diffuseness $a=0.65~$fm.
The quantity $\iol{M}_N$, which would automatically be determined 
by a self-consistent calculation, is chosen such that the
momentum sum rule for each nucleus is satisfied. Some results
are listed in Table~\ref{tab:nuclei}.

Given the radial wavefunctions, we can determine the mean values of the 
scalar and vector fields experienced by the nucleon in the state $\k$, 
that is 
\begin{align}
\label{eqn:scalar}
M_{N\k} &= \int d^3 r\ \psi^\dagger_\k (r)\,M_N(r)\,\psi_\k (r), \\
\label{eqn:vector}
V_{N\k} &= \int d^3 r\ \psi^\dagger_\k (r)\,V_N(r)\,\psi_\k (r),
\end{align}
where $M_N(r) = M_N+S_N(r)$.
Using a local density approximation in our effective quark theory,
the scalar field felt by the quarks in the nucleon can be evaluated
by determining the quark mass, $M_\k$, that gives the appropriate 
nucleon mass, $M_{N\k}$, as the solution of the quark-diquark equation.
The vector field felt by the quarks is simply one-third 
of that felt by the nucleon (i.e. $V_\k=V_{N\k}/3$). These fields
are used in the calculation of the quark distributions in the bound
nucleon. 

%===============================================================================
%###############################################################################
%===============================================================================
%\section{Medium modified quark distributions in the nucleon}

Therefore to complete our description of quark distributions in nuclei we require the
medium modified quark distributions in the bound nucleon. We give only
the briefest review of the model used to obtain these distributions 
and refer the reader to Refs.~\cite{Mineo:2003vc,Cloet:2005rt,Cloet:2005pp}, 
where the formalism is explained in considerable detail.

\begin{figure}[tbp]
\centering\includegraphics[width=\columnwidth,angle=0]{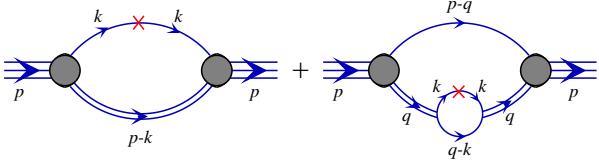}
\caption{Feynman diagrams representing the quark distributions in the nucleon. 
The single line represents the quark propagator and the double line the 
diquark $t$-matrix. The shaded oval
denotes the quark-diquark vertex function and the operator insertion has the form
$\g^+\g_5\,\d\!\lf(x - \frac{k_-}{p_-}\rg)\frac{1}{2}\lf(1 \pm \tau_z\rg)$
for the spin-dependent distribution, while $\g^+\g_5 \to \g^+$ for the spin-independent case.}
\label{fig:feydiagrams}
\end{figure}

The nucleon is described by solving the relativistic Faddeev equation including
both scalar and axial-vector diquark correlations in a confining Nambu--Jona-Lasinio
model framework. For this calculation we utilize the static approximation
for the quark exchange kernel \cite{Buck:1992wz}.
The quark distributions in the nucleon are obtained from a Feynman 
diagram calculation, where we give the relevant diagrams 
in Fig.~\ref{fig:feydiagrams}.
Medium effects are included by introducing the scalar and vector mean-fields,
obtained from Eqs.~(\ref{eqn:scalar},\ref{eqn:vector}), into the quark propagators.
Inclusion of the vector field leads to a density dependent shift in the Bjorken 
scaling variable. Fermi motion effects are included via convolution
with the smearing functions (Eq.~(\ref{eqn:fkk}) or (\ref{eqn:gkk})) 
after introducing 
the scalar field, but before the shift required by the vector field.

The vector field dependence of the quark distributions in 
the nucleon (with momentum $p^\mu$ and mean fields for 
the state $\k$) is given by
\be
\D q_{\a,\k}(x) = \frac{p_-}{p_{N-}^\k}\,
\D q_{\a 0,\k}\lf(\frac{p_-}{p_{N-}^\k}x - \frac{V_{-}^\k}{p_{N-}^\k}\rg),
\label{eqn:minshift}
\ee
where $p_{N-}^\k = p_- - 3V_-^\k$ and $\D q_{\a 0,\k}$ is 
the quark distribution without the 
vector field \cite{Mineo:2003vc}.
Here $V_-$ is the minus component of the vector field, 
$V_\mu \equiv (V_0,\vec{0}\,)$, acting on a quark.\footnote{
For the light-cone coordinates we use 
$a_\pm=\tfrac{1}{\sqrt{2}}(a_0 \pm a_3)$.}
If we now define the auxiliary quantities
\begin{align}
E_\k \equiv \ve_\k - V_{N\k}, \qquad \hat{M}_{N\k} \equiv \iol{M}_N - V_{N\k},
\end{align}
it is easy to rewrite the $\d$-function in Eq.~(\ref{eqn:gyA}) to show
\begin{multline}
\D f_{\k m}(y_A) = \frac{\iol{M}_N}{\hat{M}_{N\k}}
\D f_{0,\k m}\lf(\frac{\iol{M}_N}{\hat{M}_{N\k}}\,y_A - \frac{V_{N\k}}{\hat{M}_{N\k}}\rg),
\label{eqn:gyAshift}
\end{multline}
where the function $\D f_{0,\k m}$ has the same form as 
Eq.~(\ref{eqn:gyA}), except for
the replacements $\ve_\k \to E_\k$ and $\iol{M}_N \to \hat{M}_{N\k}$. 
Substituting Eqs.~(\ref{eqn:minshift},\ref{eqn:gyAshift}) into Eq.~(\ref{eqn:con})
and performing an analogous calculation to that found 
in Appendix C of Ref.~\cite{Mineo:2003vc}
we obtain
\begin{multline}
\D q_{\a,\k}^m(x_A) = \frac{\iol{M}_N}{\hat{M}_{N\k}} 
\D q_{\a0,\k}^m\lf(\frac{\iol{M}_N}{\hat{M}_{N\k}}\,x_A - \frac{V_\k}{\hat{M}_{N\k}}\rg),
\label{eqn:finiteshift}
\end{multline}
where the distribution, $\D q^{m}_{\a0,\k}$, is given by the convolution 
of $\D q_{\a0,\k}$ and $\D f_{0,\k m}$.
An important feature of this approach is that the number and 
momentum sum rules are satisfied from the outset.

%-------------------------------------------------------------------------------
%-------------------------------------------------------------------------------

%===============================================================================
%###############################################################################
%===============================================================================
\section{Results}
\label{sec:results}

The parameters for the quark-diquark model for the bound nucleon are discussed in 
Refs.~\cite{Cloet:2005pp,Cloet:2005rt}, so we will not repeat them. 
The new features 
presented in this paper are those associated with finite nuclei. 
In Table~\ref{tab:nuclei} 
we give values for $\iol{M}_{N}$, $M_{N\k}$ and $V_{N\k}$ obtained from the
single particle shell model. 
These values are then used in Eq.~(\ref{eqn:con})
to calculate the nuclear quark distributions.

The unpolarized EMC effect is defined as the ratio of the spin-averaged structure
function, $F_{2A}$, of a particular nucleus $A$ divided by the naive expectation. That is
\be
R_A = \frac{F_{2A}}{F_{2A}^{\text{naive}}} = \frac{F_{2A}}{Z\,F_{2p} + (A-Z)\,F_{2n}},
\label{eqn:EMC}
\ee
where $F_{2p}$ is the free proton structure function and $F_{2n}$ the free neutron
structure function\footnote{Experimental EMC ratios for $N\simeq Z$ nuclei are 
usually determined with the deuteron structure function $F_{2D}$ in the denominator. In our 
mean field model we assume $F_{2D}\simeq F_{2p}+F_{2n}$. We therefore anticipate 
deuteron binding corrections of a few percent to our EMC ratios for $x \gtrsim 0.5$,
when comparing with experimental data.}.  
In the limit of no Fermi motion and no medium effects of any kind, this ratio is unity. 
An equivalent EMC ratio can also be defined for the $K=0$ multipole.

The polarized EMC effect is defined by an analogous ratio, 
which is the spin-dependent 
structure function for a particular nucleus with helicity $H$,
divided by the naive expectation, that is
\be
R_{As}^{JH} =\frac{g_{1A}^{JH}}{g^{JH}_{1A,\text{naive}}}
=\frac{g_{1A}^{JH}}{P^{JH}_{p}\,g_{1p} + P^{JH}_{n}\,g_{1n}}.
\label{eqn:pEMC}
\ee
Here $g_{1p}$ and $g_{1n}$ are the free nucleon structure functions
and $P^{JH}_{p(n)}$ is the polarization of the protons (neutrons)
in the nucleus with helicity $H$, defined by
\be
P^{JH}_\a = \la J,H \lvert2\,\hat{S}^\a_z\rvert J,H\ra, \qquad \a \in \lf(p,n\rg),
\label{eqn:pol}
\ee
where $\hat{S}_z^\a$ is the total spin operator for protons or neutrons.
From an experimental standpoint one should simply use the best estimates
of the polarization factors available in the literature.
In this work we use the polarization factors obtained from the
nonrelativistic limit of Eq.~(\ref{eqn:pol}), which differ from the relativistic 
values calculated within our model by less than 2\%.
If only a single valence nucleon or nucleon-hole contributes to
the nuclear polarization, then in the nonrelativistic limit the polarization 
factor is simply given by
\be
P^{JH}_\a = \pm\frac{2\,H}{2\ell + 1},
\label{eqn:polv}
\ee
where $\ell$ is the orbital angular momentum and the $\pm$ refers to the 
cases $J = \ell \pm \tfrac{1}{2}$.

%===============================================================================
\begin{table}[tbp]
\addtolength{\tabcolsep}{3.0pt}
\addtolength{\extrarowheight}{3.0pt}
\begin{tabular}{c|c|cccc|cccc}
\hline\hline
          & $\iol{M}_{\!N}$ & \multicolumn{4}{c|}{$M_{N\k}$} & \multicolumn{4}{c}{$V_{N\k}$} \\
\cline{3-10}
          &                 & -1  & -2  & 1    & -3          &  -1 &  -2 &  1  & -3 \\
\hline
$^7$Li    & 933             & 811 & 856 & --   & --          &  89 &  58 &  -- & -- \\
$^{11}$B  & 931             & 793 & 829 & --   & --          & 101 &  76 &  -- & -- \\
$^{15}$N  & 929             & 785 & 815 & 815  & --          & 106 &  86 &  86 & -- \\
$^{27}$Al & 930             & 771 & 794 & 793  & 820         & 115 & 101 & 101 & 82 \\
\hline\hline
\end{tabular}
\caption{All quantities are in MeV. The labels $-$1, $-$2, 1, $-$3 refer 
to the Dirac
quantum number $\k$, where $\lvert\k\rvert = j+\tfrac{1}{2}$.}
\label{tab:nuclei}   
\end{table}
%===============================================================================

The  polarized EMC ratio can also be defined for the $K=1$ multipole structure 
function and has the form
\be
R_{As}^{(J1)} =\frac{g_{1A}^{(J1)}}{P^{(J1)}_{p}\,g_{1p} + P^{(J1)}_{n}\,g_{1n}},
\label{eqn:pEMCMult}
\ee
where $P^{(J1)}_\a$ is the reduced matrix element
\be
P^{(J1)}_\a = \la J\lvert\lvert\,2\,\hat{S}^\a\,\rvert\rvert J\ra
=\sqrt{\frac{(2J+1)(2J+2)}{6J}}\,P_\a^{JJ}.
\label{eqn:polmulti}
\ee

Because the spin structure function $g_{1n}$ is smaller than $g_{1p}$ and remains poorly known,
especially at large $x$, it is clear from Eqs.~(\ref{eqn:pEMC},\ref{eqn:pEMCMult})
that to determine the polarized EMC effect it is necessary to
choose nuclei where $\lvert P_n\rvert \ll \lvert P_p\rvert$.
There is also an upper limit on the
mass number of nuclei 
that can be readily used to measure the polarized EMC effect,
because for spin cross-sections the valence nucleons dominate 
and hence $g_{1A}$ is 
suppressed by approximately $1/A$ relative to $F_{2A}$, where all nucleons contribute.

The best candidates are nuclei with a single valence proton 
or proton-hole, for example 
the stable nuclei $^{11}$B, $^{15}$N and $^{27}$Al. Another
good choice is $^7$Li, where the nuclear polarization is largely dominated
by the valence proton. Extensive studies of $^7$Li, beginning in the 60's
with the shell model \cite{Cohen:1965qa}, to 
modern state of the art Quantum Monte 
Carlo calculations \cite{Pudliner:1997ck},  
consistently find $P_p \simeq 0.86-0.88$.
The Quantum Monte Carlo result for the neutron polarization 
is $P_n \simeq -0.04$.

First we discuss the nuclear quark distributions, focusing 
on $^7$Li as its treatment is 
a little more involved compared with the other nuclei, because there 
are three valence nucleons coupled to $J=3/2$ and $T=1/2$.
Using the $H=3/2$ shell model wavefunction found 
in Refs.~\cite{Landau:1977,Guzey:1999rq},
when evaluating the spin-independent analogue of 
Eq.~(\ref{eqn:con}) for the $u$-quark 
distribution in $^7$Li we obtain
\begin{multline}
u_A^{3/2\,3/2}(x_A) = 2\lf[u_{p,-1}^{1/2}(x_A) + d_{p,-1}^{1/2}(x_A)\rg] \\
+ \frac{1}{15}\biggl[ 13\,u_{p,-2}^{3/2}(x_A) + 20\,d_{p,-2}^{3/2}(x_A) \\
+ 2u_{p,-2}^{1/2}(x_A) + 10\,d_{p,-2}^{1/2}(x_A) \biggr],
\label{eqn:upLi7}
\end{multline}
where we have used charge symmetry to relate $u_n \leftrightarrow d_p$.
The spin-dependent distribution has the form
\begin{align}
\D u_A^{\,3/2\,3/2}(x_A) &= \frac{1}{15}\lf[ 13\,\D u_{p,-2}^{3/2}(x_A) + 2\,\D d_{p,-2}^{3/2}(x_A) \rg].
\label{eqn:DupLi7}
\end{align}
Similar expressions hold for the $H=1/2$ and $d$-quark distributions. 
With this wavefunction the $^7$Li 
polarization factors are $P_p^{JH} = \tfrac{2H}{3}\tfrac{13}{15}$
and $P_n^{JH} = \tfrac{2H}{3}\tfrac{2}{15}$.
For the other nuclei the situation is simpler as we 
make the approximation that 
the nuclear spin is carried solely by the valence proton-hole. 

\begin{figure}[tbp]
\centering\includegraphics[width=\columnwidth,clip=true,angle=0]{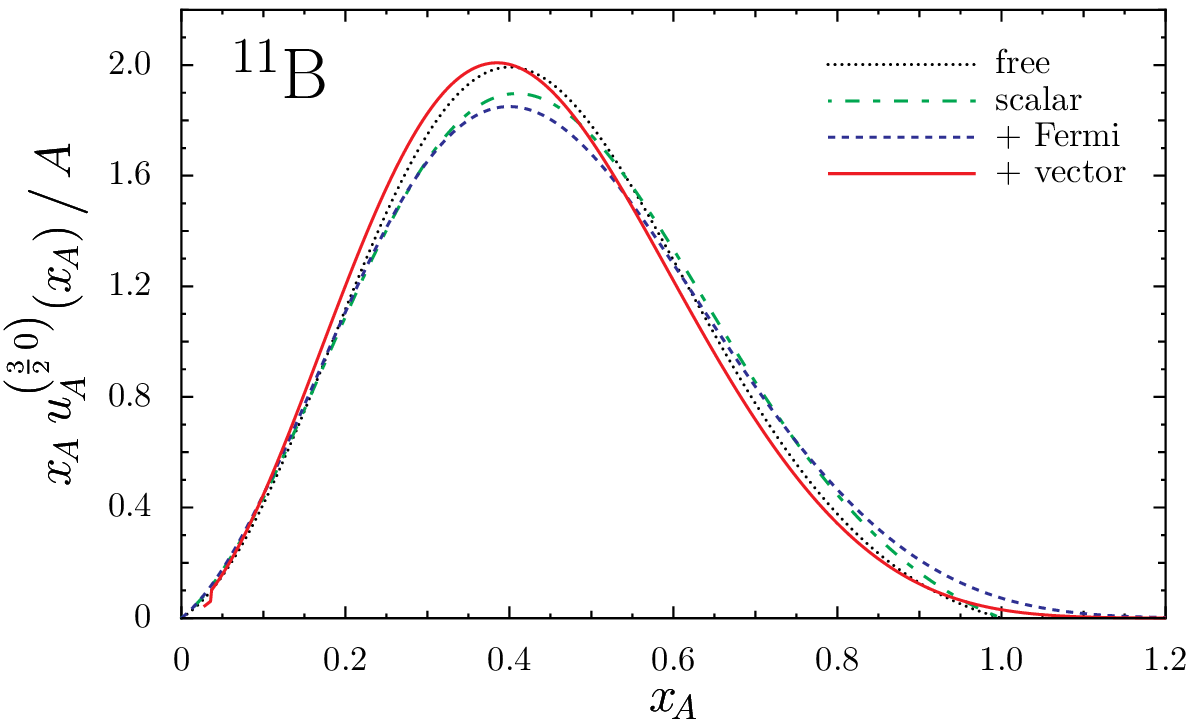}
\vskip -1em
\caption{The first spin-independent multipole (K=0) $u$-quark distribution in $^{11}$B
(at $Q^2=Q_0^2$).}
\vskip 1em
\label{fig:1a}
\centering\includegraphics[width=\columnwidth,clip=true,angle=0]{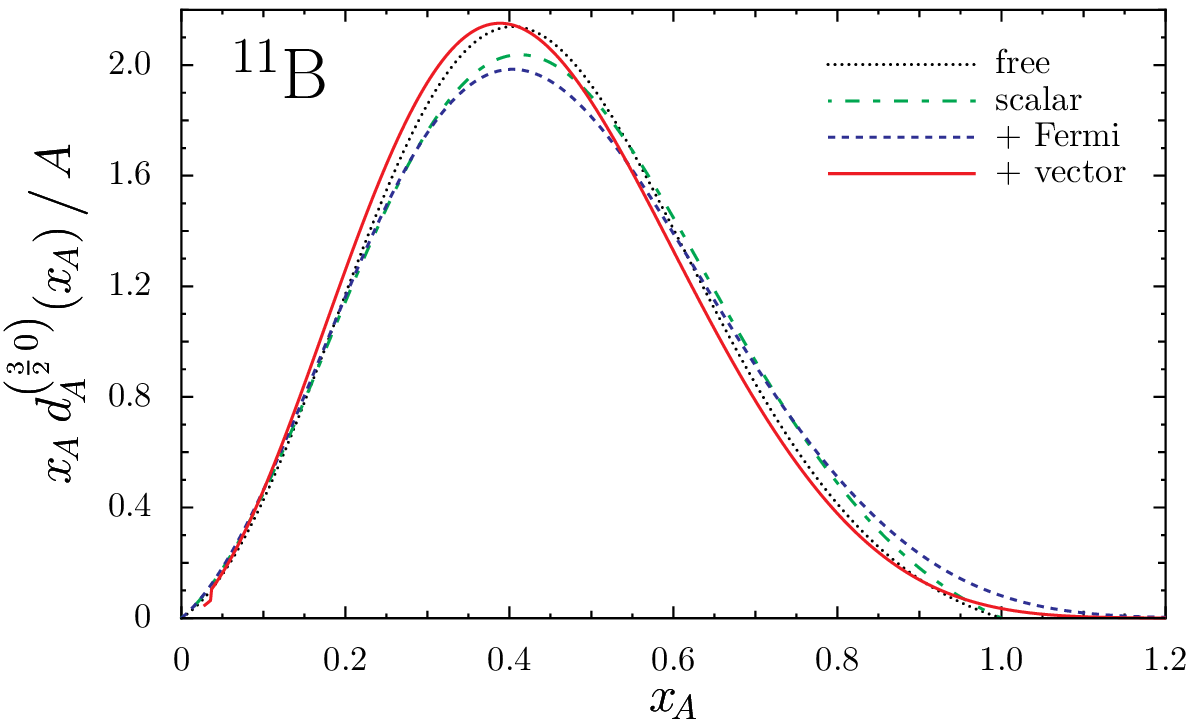}
\vskip -1em
\caption{The first spin-independent multipole (K=0) $d$-quark distribution in $^{11}$B
(at $Q^2=Q_0^2$).}
\vskip 1em
\label{fig:2a}
\end{figure}

\begin{figure}[tbp]
\vskip -0.3em
\centering\includegraphics[width=\columnwidth,clip=true,angle=0]{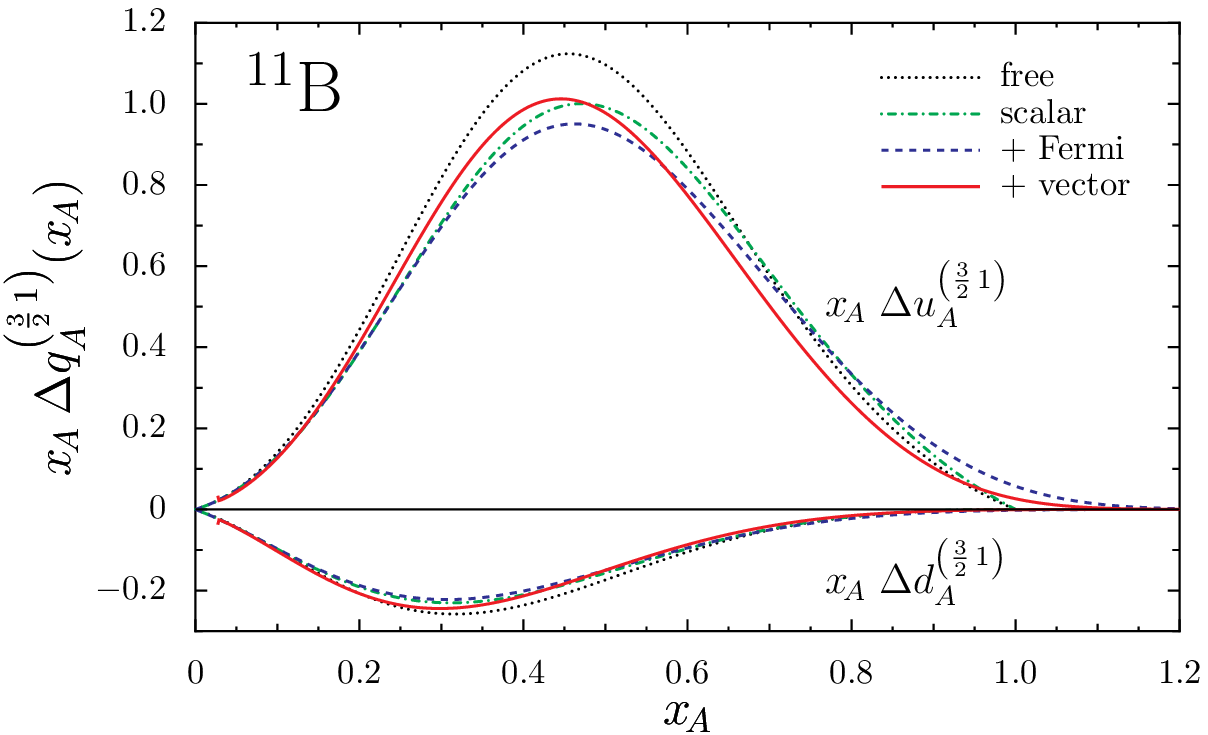}
\vskip -1em
\caption{The first spin-dependent multipole (K=1) $u$- and $d$-quark distributions in $^{11}$B
(at $Q^2=Q_0^2$).}
\vskip 1em
\label{fig:3a}
\centering\includegraphics[width=\columnwidth,clip=true,angle=0]{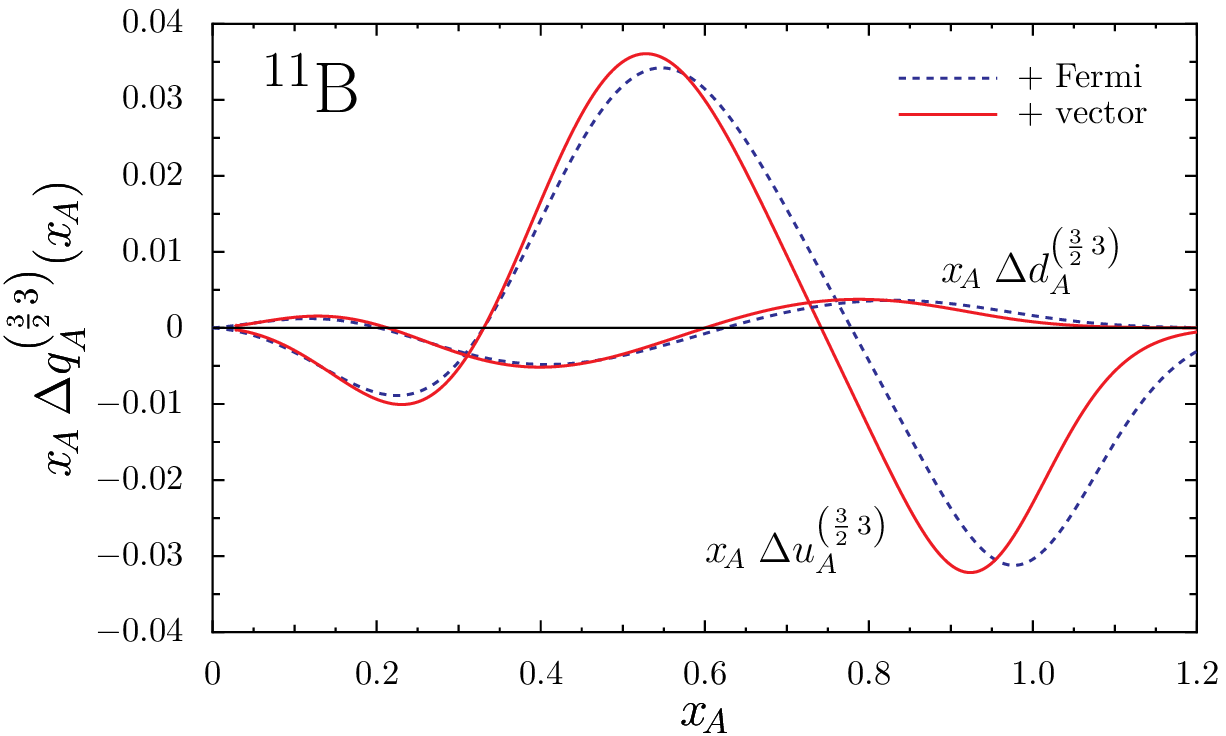}
\vskip -1em
\caption{The second spin-dependent multipole (K=3) $u$- and $d$-quark distributions in $^{11}$B
(at $Q^2=Q_0^2$).}
\vskip 1em
\label{fig:4a}
\end{figure}

In Figs.~\ref{fig:1a}-\ref{fig:4a} we show the leading multipole quark distributions for
$^{11}$B, together with the next-to-leading $K=3$ multipole for the spin-dependent case,
at the model scale of $Q_0^2= 0.16~$GeV$^2$ \cite{Cloet:2005pp}.
The other nuclear quark distributions are similar, so we will not show them. 
The dotted line is the result 
without Fermi motion and medium effects,
and is obtained from expressions like Eq.~(\ref{eqn:con}) by 
replacing each smearing function with a 
delta function (multiplied by the polarization factor
for the spin-dependent case) and 
using the free results for the $u$- and $d$-quark distributions
in the nucleon. 
The dot-dashed line includes the effect of the scalar field, and
the dashed curve also incorporates Fermi motion, which is the result after convolution
with the appropriate nucleon distribution (Eqs.~(\ref{eqn:fkk},\ref{eqn:gkk})). 
The complete in-medium distribution is given by
the solid line and is the result obtained after 
also shifting the scaling variable using
Eq.~(\ref{eqn:finiteshift}). 

For the spin-independent distributions all nucleons contribute. Therefore,
in Figs.~\ref{fig:1a} and \ref{fig:2a} we see that the $u$- and $d$-quark
distributions are very similar. For the spin-dependent case (see Fig.~\ref{fig:3a}) only the 
valence proton-hole contributes. Hence the distributions resemble those of the proton.
We find that higher multipole distributions 
are greatly suppressed relative to the leading results, see for example
the $K=3$ distribution in Fig.~\ref{fig:4a}.
The spin-independent $K=2$ multipole is an order of magnitude smaller again
and reflects the very weak helicity dependence of 
the $F_{2A}^{JH}$ structure functions. 
This weak helicity dependence arises because the spin-zero core is the dominant
contribution to $F_{2A}^{JH}$, and changes in $H$ simply reflect 
different spin orientations of the valence nucleon(s). 

%===============================================================================

%=============================================================================== 

The main features of the medium effects displayed 
in Figs.~\ref{fig:1a}\,--\,\ref{fig:3a}
are similar to those found in our earlier 
nuclear matter calculation \cite{Cloet:2005rt}.
The spin-independent distributions are quenched at large $x$ and 
enhanced for small $x$, whereas the spin-dependent distributions 
are quenched for all $x$.

The nuclear spin sum, $\Sigma^{(A)}$, and axial coupling, $g_A^{(A)}$, 
contain information on both 
nuclear and quark effects and are simply given by
\begin{align}
\Sigma^{(A)}  &= \D u_A + \D d_A \equiv \Sigma\,\lf(P_p+P_n\rg), \\
g_A^{(A)} &= \D u_A - \D d_A \equiv g_A\,\lf(P_p-P_n\rg),
\end{align}
where $\D q_A$ represents the first moment of $\D q_A^{JJ}$ and $\Sigma$, $g_A$
are the medium modified nucleon quantities, defined by dividing out the 
non-relativistic isoscalar and isovector polarization factors for $H=J$.
We find that $\Sigma$ and $g_A$ are both suppressed in-medium
relative to the free values, as summarized Table~\ref{tab:2}.
This decrease of $g_A$ in-medium is in agreement with the well known 
nuclear $\beta$-decay studies which, after taking into account the nuclear
structure effects, require a quenching of $g_A$ to achieve agreement with 
empirical data.\footnote{The required quenching factors can be seen, 
for example, by comparing the experimental and calculated Gamow-Teller 
matrix elements given in Refs.~\cite{Brown:1983,Suzuki:2003fw}.}

%===============================================================================
\begin{table}[tp]
\addtolength{\tabcolsep}{10.0pt}
\addtolength{\extrarowheight}{1.0pt}
\begin{tabular}{c|cccccccccc}
\hline\hline
           &  $\D u$ & $\D d$ & $\Sigma$  & $g_A$ \\[0.2ex]
\hline\\[-1.9ex]
$p$        & 0.97   & -0.30  & 0.67  & 1.267   \\
$^7$Li     & 0.91   & -0.29  & 0.62  & 1.19\,~ \\
$^{11}$B   & 0.88   & -0.28  & 0.60  & 1.16\,~ \\
$^{15}$N   & 0.87   & -0.28  & 0.59  & 1.15\,~ \\
$^{27}$Al  & 0.87   & -0.28  & 0.59  & 1.15\,~ \\
Nucl. Matter & 0.74   & -0.25  & 0.49  & 0.99\,~ \\
\hline\hline
\end{tabular}
\caption{Results for the first moment of the in-medium 
quark distributions in the bound proton and the resulting spin sum and 
nucleon axial charge. It is clear that the moments tend 
to decrease with increasing $A$.}   
\label{tab:2}
\end{table}
%===============================================================================

\begin{figure}[tbp]
\centering\includegraphics[width=\columnwidth,angle=0]{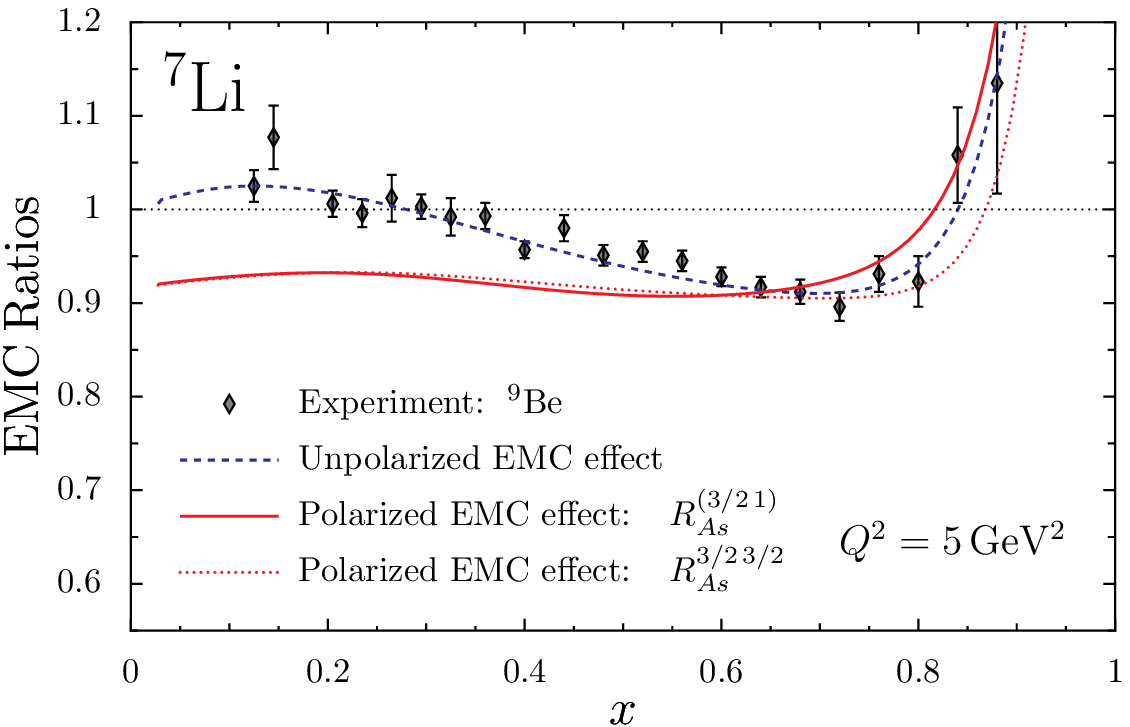}
\vskip -1em
\caption{The EMC and polarized EMC effect in $^7$Li. 
The empirical data is from Ref.~\cite{Gomez:1993ri}.}
\vskip 1em
\label{fig:1}
\centering\includegraphics[width=\columnwidth,angle=0]{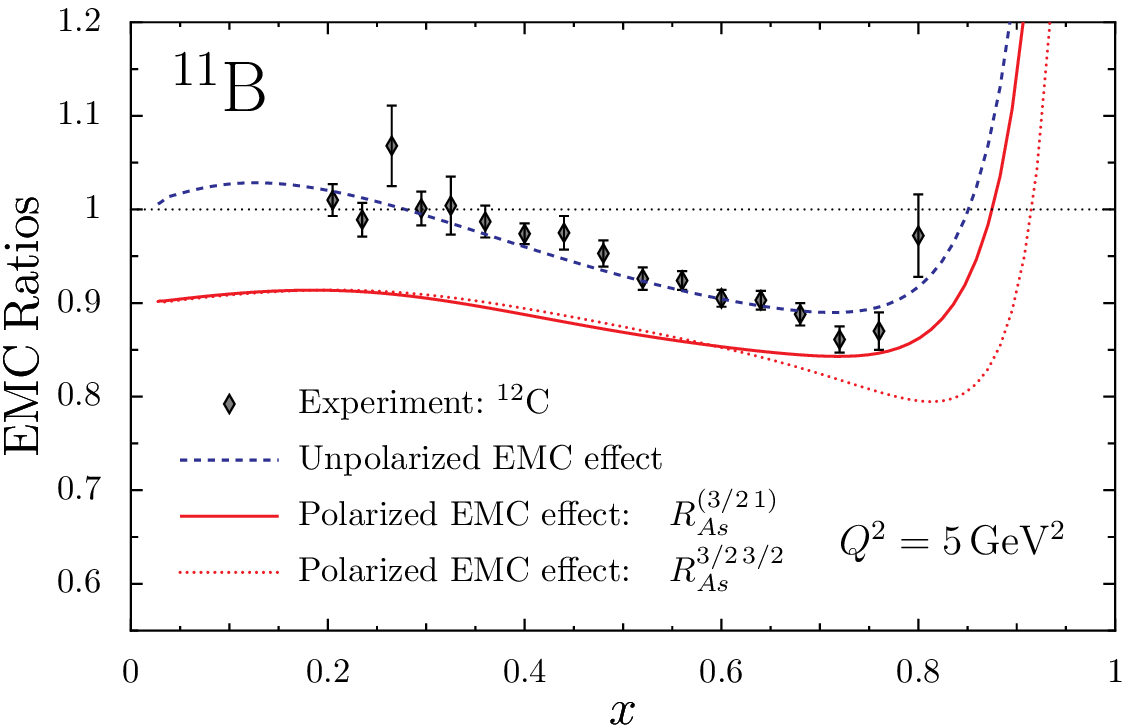}
\vskip -1em
\caption{The EMC and polarized EMC effect in $^{11}$B. 
The empirical data is from Ref.~\cite{Gomez:1993ri}.}
\vskip 1em
\label{fig:2}
\end{figure}

\begin{figure}[!t]
\centering\includegraphics[width=\columnwidth,angle=0]{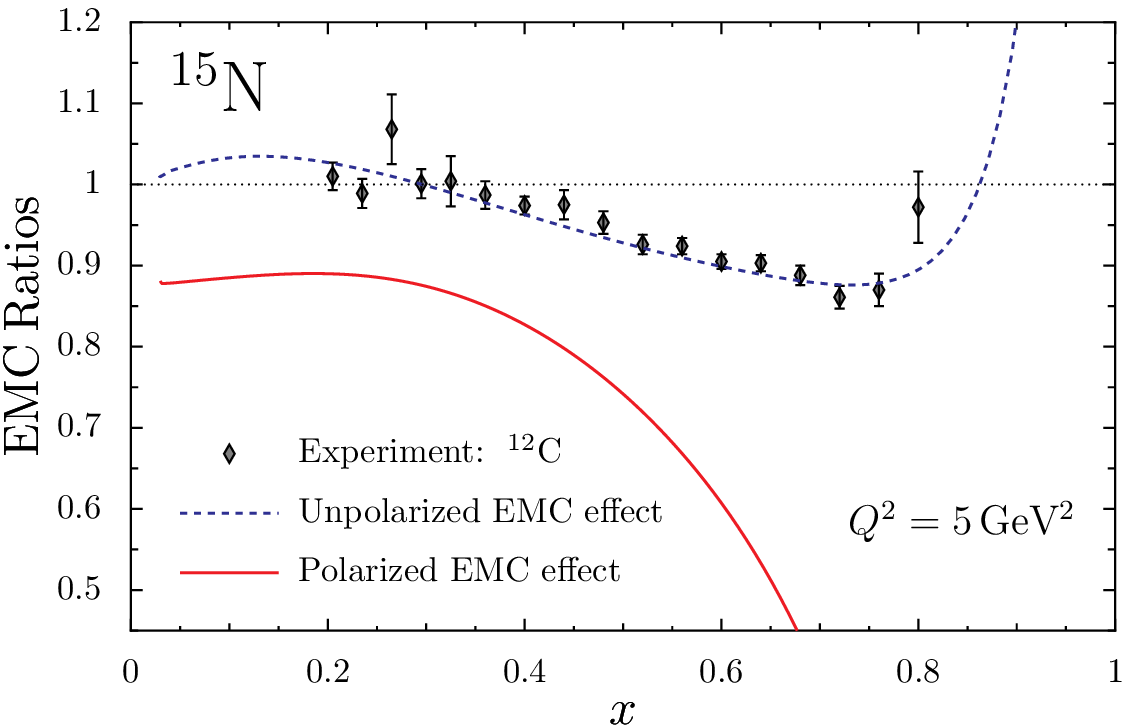}
\vskip -1em
\caption{The EMC and polarized EMC effect in $^{15}$N. 
The empirical data is from Ref.~\cite{Gomez:1993ri}.}
\vskip 1em
\label{fig:3}
\centering\includegraphics[width=\columnwidth,angle=0]{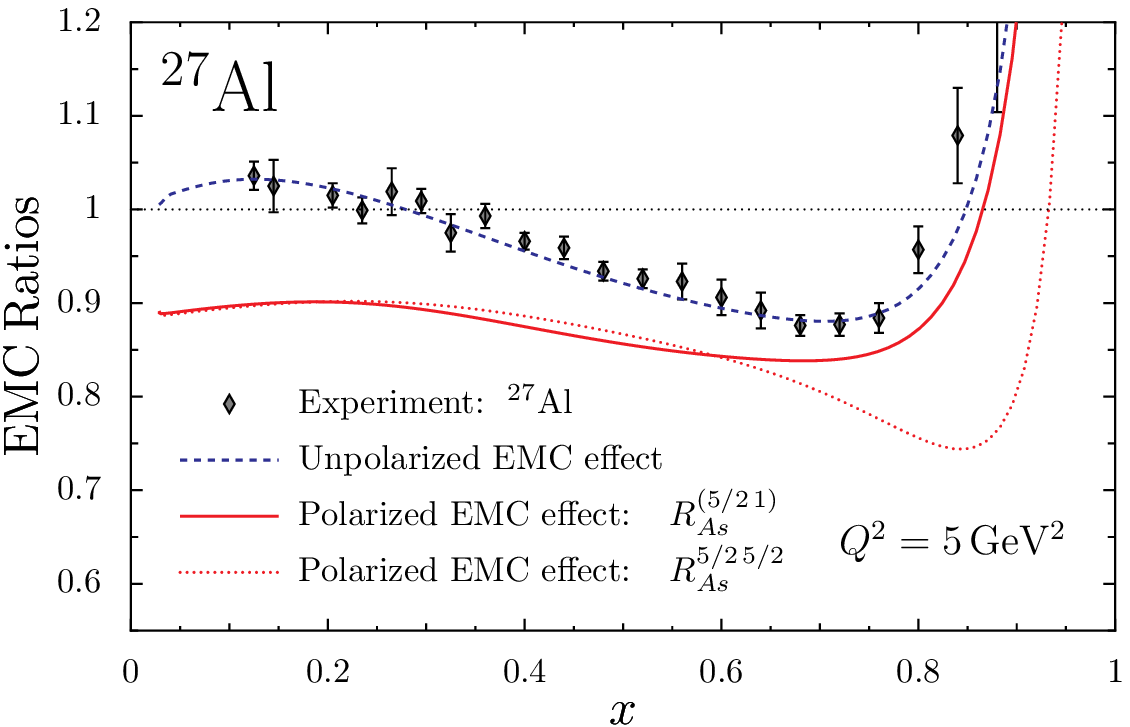}
\vskip -1em
\caption{The EMC and polarized EMC effect in $^{27}$Al. 
The empirical data is from Ref.~\cite{Gomez:1993ri}.}
\vskip 1em
\label{fig:4}
\end{figure}

In Figs.~\ref{fig:1}\,--\,\ref{fig:4} we give results for the EMC and polarized 
EMC effect in $^7$Li, $^{11}$B, $^{15}$N and $^{27}$Al at $Q^2 = 5\,$GeV$^2$. 
The dashed line is the unpolarized EMC effect, the solid line is the $K=1$ polarized 
EMC effect and the dotted line is the $M=J$ polarized EMC result
(c.f. Eqs.~(\ref{eqn:pEMCMult}) and (\ref{eqn:pEMC}), respectively).
For the unpolarized EMC effect the results agree very 
well with the experimental data taken 
from Ref.~\cite{Gomez:1993ri}, where importantly 
we obtain the correct $A$-dependence. 

%===============================================================================

%===============================================================================

Consistent with previous nuclear matter studies, 
we find that the polarized EMC effect 
is larger than the unpolarized case, with the exception 
of the multipole result for $^7$Li at $x \gtrsim 0.6$.
Based on the wavefunction in Ref.~\cite{Guzey:1999rq}
the neutrons give a small contribution to the polarization. 
To test the dependence on the neutron
polarization we also coupled the two neutrons to spin-zero, 
so that $P_n^{3/2\,3/2}=0$, which is
closer to the Quantum Monte Carlo result of $-0.04\,$\cite{Pudliner:1997ck}. We find that these results
are very similar to those in Fig.~\ref{fig:1}.

The unusual shape for the $^{15}$N polarized EMC result 
is because our full result for $g_{1A}^{1/2\,1/2}$ changes
sign at $x \simeq 0.8$, and hence the ratio must go to zero at this point.
The origin of this sign change is the nucleon 
$p_{1/2}$ smearing function, which becomes positive for large $y_A$.
This result suggests $^{15}$N
may not be a good candidate with which to study nucleon medium modifications. 
The $^{11}$B and $^{27}$Al results resemble those obtained for 
nuclear matter \cite{Cloet:2005rt}, where we find a polarized EMC 
effect roughly twice that of the unpolarized case.

\section{Conclusion}

Using a relativistic formalism, where the quarks in the bound 
nucleons respond to the nuclear environment, we calculated the quark
distributions and structure functions of $^7$Li, $^{11}$B, $^{15}$N
and $^{27}$Al. For a spin-$J$ target there are $2J+1$ 
independent quark distributions or structure functions in the Bjorken limit. 
For example, $^{27}$Al therefore has six structure functions, however we find
that the higher multipoles are suppressed relative to the leading result by
at least an order of magnitude.

We were readily able to describe the EMC effect in these nuclei,
and importantly obtained the correct $A$-dependence.
Although we do not show the results, we also determined the 
EMC ratio for $^{12}$C, $^{16}$O and $^{28}$Si and found results
very similar to their $A-1$ neighbours.
In Eq.~(\ref{eqn:pEMC}) we define the polarized EMC ratio in nuclei. 
This ratio is such that in the extreme nonrelativistic limit, with 
no medium modifications, it is unity. 
The results for the polarized EMC effect in nuclei corroborate 
earlier nuclear matter \cite{Cloet:2005rt,Smith:2005ra}, light
nuclei \cite{Steffens:1998rw} and small $x$ \cite{Guzey:1999rq}
studies that found large medium modifications to the spin structure
function relative to the unpolarized case. 
In particular, we find that the fraction of the spin of the nucleon
carried by the quarks is decreased in nuclei (see Table~\ref{tab:2}).
If this result is confirmed experimentally, it would give important insights
into in-medium quark dynamics and help quantify the role of quark degrees of
freedom in the nuclear environment.

\vspace*{-0.6em}
\section*{Acknowledgments} 
\vspace*{-0.6em}

W.B. wishes to thank S. Kumano, T. Suzuki (Nihon University) and B. A. Brown for 
discussions.  We also thank S. Kumano for the QCD evolution code \cite{Miyama:1995bd,Hirai:1997gb}.
This work was supported by the Australian Research Council and DOE 
contract DE-AC05-84150,
under which JSA operates Jefferson Lab, and by the Grant in Aid for Scientific
Research of the Japanese Ministry of Education, Culture, Sports, Science and
Technology, Project No. C2-16540267.

\end{document}